\documentclass[aps,pra,reprint,superscriptaddress,letterpaper,showpacs]{revtex4-1}
\usepackage[english]{babel}
\usepackage{ucs}
\usepackage[utf8x]{inputenc}
\usepackage{amsmath}
\usepackage{amsfonts}
\usepackage{amssymb}
\usepackage{graphicx}
\usepackage{bm}
\usepackage{bbm}
\usepackage{braket}

\begin{document}

\title{Spectator Errors in Tunable Coupling Architectures}

\author{D.\ M.\ Zajac}
\thanks{These authors contributed equally}
\affiliation{IBM Quantum, IBM T.J.\ Watson Research Center, Yorktown Heights, NY 10598, USA}
\author{J.\ Stehlik}
\thanks{These authors contributed equally}
\affiliation{IBM Quantum, IBM T.J.\ Watson Research Center, Yorktown Heights, NY 10598, USA}
\author{D.\ L.\ Underwood}
\affiliation{IBM Quantum, IBM T.J.\ Watson Research Center, Yorktown Heights, NY 10598, USA}
\author{T.\ Phung}
\affiliation{IBM Quantum, IBM Almaden Research Center, San Jose, CA 95120, USA}
\author{J. Blair}
\affiliation{IBM Quantum, IBM T.J.\ Watson Research Center, Yorktown Heights, NY 10598, USA}
\author{S.\ Carnevale}
\affiliation{IBM Quantum, IBM T.J.\ Watson Research Center, Yorktown Heights, NY 10598, USA}
\author{D.\ Klaus}
\affiliation{IBM Quantum, IBM T.J.\ Watson Research Center, Yorktown Heights, NY 10598, USA}
\author{G.\ A.\ Keefe}
\affiliation{IBM Quantum, IBM T.J.\ Watson Research Center, Yorktown Heights, NY 10598, USA}
\author{A.\ Carniol}
\affiliation{IBM Quantum, IBM T.J.\ Watson Research Center, Yorktown Heights, NY 10598, USA}
\author{M.\ Kumph}
\affiliation{IBM Quantum, IBM T.J.\ Watson Research Center, Yorktown Heights, NY 10598, USA}
\author{Matthias Steffen}
\affiliation{IBM Quantum, IBM T.J.\ Watson Research Center, Yorktown Heights, NY 10598, USA}
\author{O.\ E.\ Dial}
\affiliation{IBM Quantum, IBM T.J.\ Watson Research Center, Yorktown Heights, NY 10598, USA}

\begin{abstract}
The addition of tunable couplers to superconducting quantum architectures offers significant advantages for scaling compared to fixed coupling approaches.  In principle, tunable couplers allow for exact cancellation of qubit-qubit coupling through the interference of two parallel coupling pathways between qubits. However, stray microwave couplings can introduce additional pathways which complicate the interference effect. Here we investigate the primary spectator induced errors of the bus below qubit (BBQ) architecture in a six qubit device. We identify the key design parameters which inhibit ideal cancellation and demonstrate that dynamic cancellation pulses can further mitigate spectator errors.
\end{abstract}

\maketitle

One of the crucial assumptions of quantum error correction is that errors are uncorrelated beyond the extent of pairwise errors during two-qubit gate operations \cite{Gutierrez2016s, Greenbaum2017,Bravyi2018,Beale2018}. However, the presence of residual couplings between qubits can give rise to coherent many-qubit interactions, generally referred to as spectator errors. Many different qubit architectures exhibit increased gate errors with increasing system size or during simultaneous gate operations \cite{Mckay2019,Rudinger2019,Leung2018,Krinner2020}, even in cases where tunable coupling has been utilized \cite{chen2021}. Given the prevalence of these errors, and their significance for fault tolerance, studying the mechanisms which give rise to spectator errors is of critical importance to scaling up any qubit architecture.

An attractive approach to minimizing spectator errors is to use tunable couplers to turn off the interactions between qubits, and only turn on coupling when a two-qubit gate is being implemented. While the basic operation of these devices has been demonstrated in a number of works \cite{Stehlik2021,Yan18,Mundada19,Foxen20,Collodo2020,PhysRevApplied.6.064007,Sung2020,Negirneac2020}, the extent of the cancellation achieved between neighbors has not been thoroughly investigated in large devices. Here we study spectator errors in the bus below qubit (BBQ) architecture in a six qubit device \cite{Stehlik2021}. While we are able to achieve two-qubit gate fidelities comparable to those observed in isolated two-qubit experiments, we find significant non-idealities in the multi-qubit operation of the tunable couplers. Specifically we find that the coupler flux bias which is optimal for single-qubit operation can be different from the bias which cancels spectator errors for two-qubit gates, and that the cancellation can be conditional on the state of the qubits. We also identify the device parameters which give rise to these effects, and show that dynamic biasing of the couplers can be used to further minimize residual couplings during parallel two-qubit gate operations.

\begin{figure}
	\centering
	\includegraphics[width=1\columnwidth]{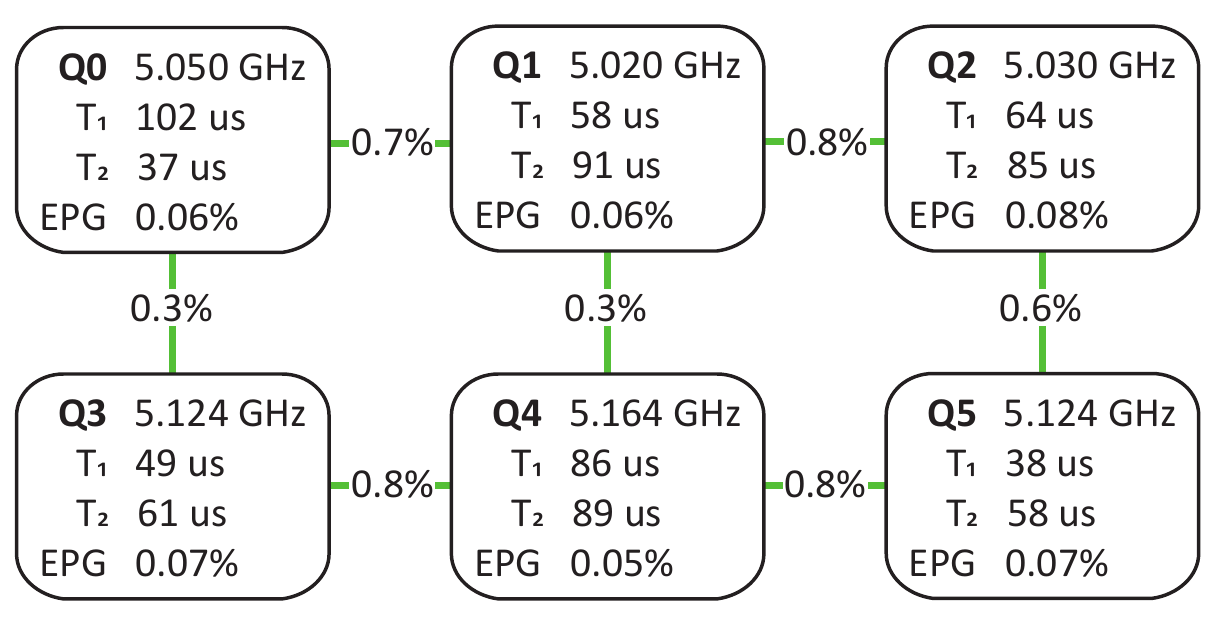}
	\caption{Six qubit BBQ device on a square lattice. With a qubit-coupler coupling strength of 80 MHz we achieve an average two-qubit gate length of 80 ns. The simultaneously measured single-qubit error-per-gate (EPG) is labelled on each qubit along with the qubit frequencies and coherences for that bias condition, and the individually optimized EPG of each two-qubit CZ gate is labelled between each pair of qubits.}
	\label{fig:device}
\end{figure}

The device used in this study, shown in Fig.~\ref{fig:device} consists of six qubits on a square lattice, with seven tunable couplers. The detunings between neighboring qubits fall in the straddling regime where the detuning is less than an anharmonicity ($\sim 240$ MHz), and the nominal coupling between qubits and couplers is 80 MHz. The relatively strong coupling between qubits and couplers combined with small detunings make this device ideal for studying spectator induced errors. By tuning the couplers to regions of low ZZ, we are able to simultaneously benchmark all single-qubit gates, achieving an average error-per-gate (EPG) of 0.064\%. However, with the couplers biased in this way we find that two-qubit gate performance is significantly worse than expected based on isolated two-qubit experiments \cite{Stehlik2021}, as we will explore in the rest of this paper. By optimizing the coupler biases for each two-qubit CZ gate separately we achieve individually benchmarked EPGs ranging from 0.3\% to 0.8\%. 

\begin{figure}%
	\centering
	\includegraphics[width=1\columnwidth]{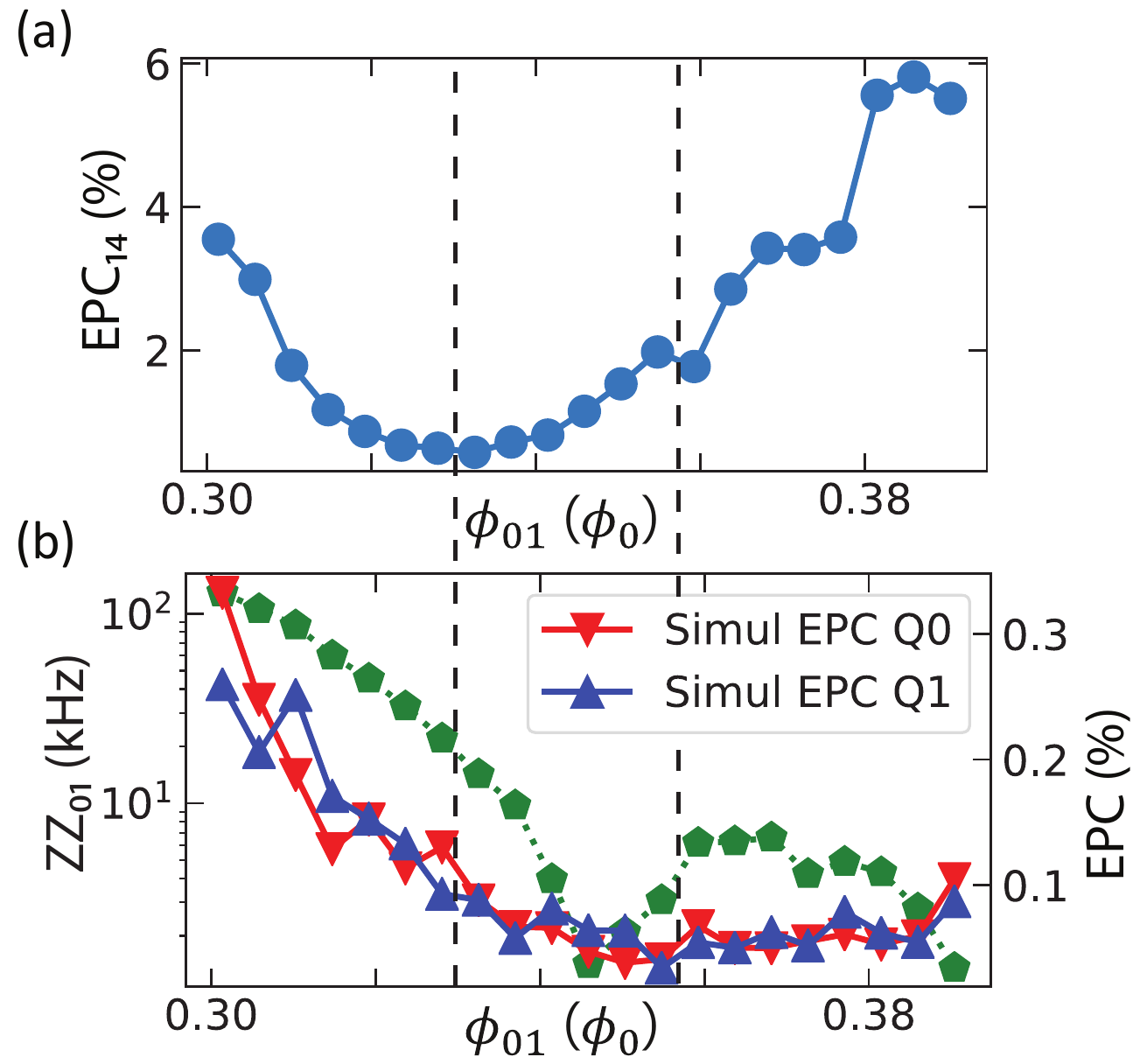}
	\caption{(a) The Q1-Q4 two-qubit error-per-Clifford(EPC) is plotted as a function of the 0-1 coupler bias $\phi_{01}$ which modulates Q0's influence as a spectator qubit. The optimal bias $\phi_{01}\sim 0.33$ $\phi_0$ is marked by a vertical dashed line. The static $ZZ$ and simultaneously measured single-qubit EPC for Q0 and Q1 is plotted on the same horizontal axis in (b), with the optimal bias point for single-qubit operations near $\phi_{01}\sim 0.36$ $\phi_0$.}
	\label{fig:rb}
\end{figure}

To demonstrate the influence of spectator errors in this device we focus on qubits 0, 1, 3 and 4. In Fig.~\ref{fig:rb} (a) we plot the fidelity of the 1-4 gate as a function of the 0-1 coupler bias. Ideally the fidelity of the 1-4 gate would be unaffected by the 0-1 coupler. However, the strong dependence of the 1-4 fidelity on the 0-1 coupler flux bias indicates the presence of spectator induced errors. Even more surprisingly, when we look at the single-qubit performance of qubits 0 and 1, we find that a different bias point is optimal for simultaneous single-qubit operations. Figure~\ref{fig:rb} (b) shows the measured ZZ interaction between qubits 0 and 1 as well as the simultaneously measured single-qubit error-per-Clifford (EPC) as a function of the 0-1 bias. Comparing these two datasets, we see that the optimal bias for single-qubit operations is near 0.36 $\phi_0$  while spectator errors on the 1-4 gate are minimized at a bias of 0.33 $\phi_0$, as indicated by the vertical dashed lines.

\begin{figure}%
	\centering
	\includegraphics[width=1\columnwidth]{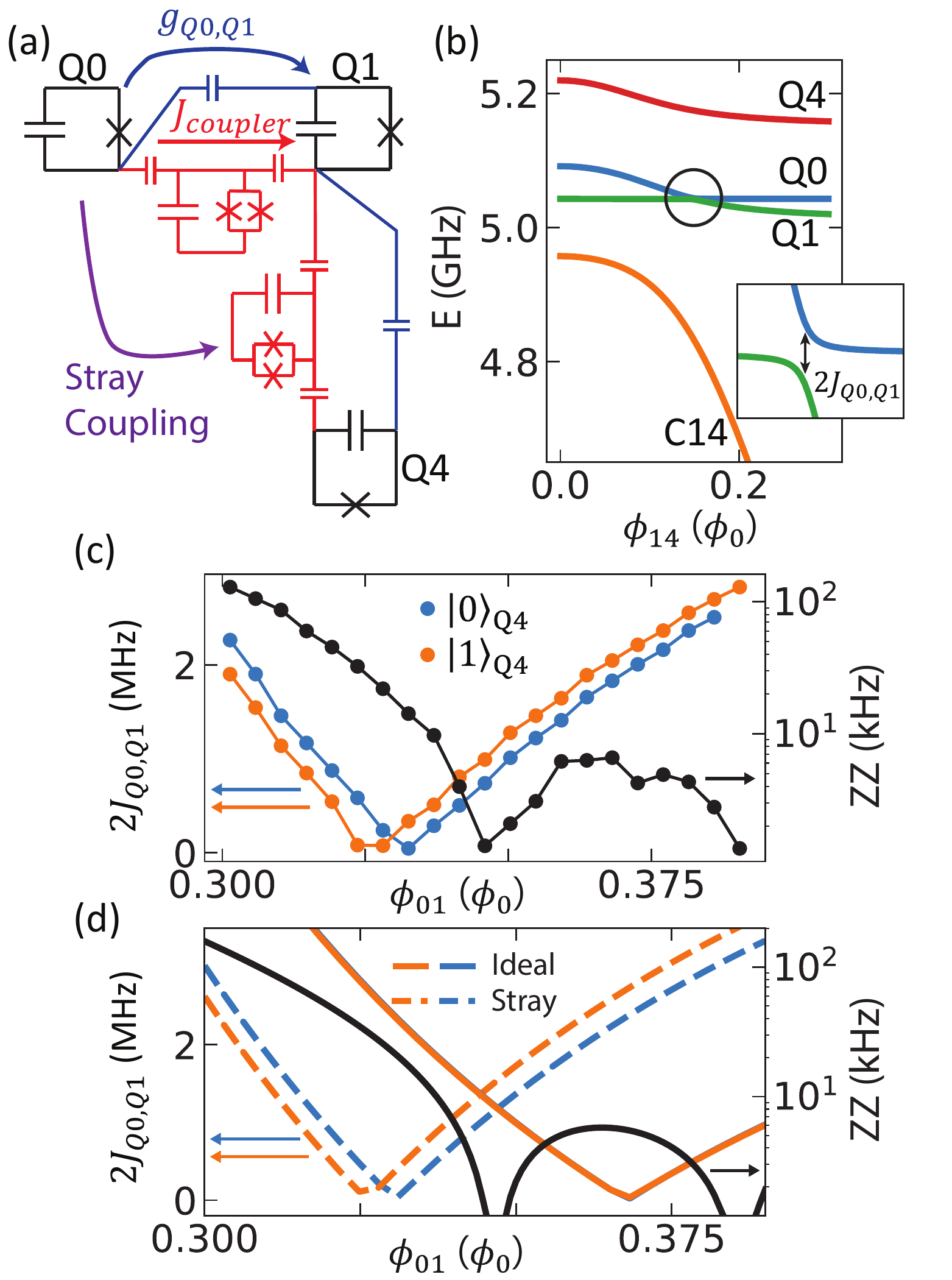}
	\caption{(a) Schematic of qubits 0, 1, and 4. Q0's coupling to Q1 and Q4 is dominated by a direct capacitive coupling $g_\text{Q0,Q1}$, a coupling mediated by C01, $J_\text{coupler}$, and an unintended, stray coupling to C14. (b) As the bias $\phi_{14}$ is pulsed to drive a Q1-Q4 two-qubit gate, Q1 traverses an avoided level crossing (inset) with Q0 whose size is determined by the net coupling from Q0 to Q1. (c) The size of the gap $2J_{\text{Q0,Q1}}$, measured via the swap rate between Q0 and Q1 is plotted, along with $ZZ$, as a function of $\phi_{01}$ for Q4 in the ground state (blue) and excited state (orange). The expected swap rates for Q4 in the ground state (blue) and excited state (orange) are plotted in (d) based on the ideal Hamiltonian (solid) and the Hamiltonian with stray couplings present (dashed). }
	\label{fig:gaps}
\end{figure}

We can gain further insight into the discrepancy of the optimal bias points for single-qubit operation and spectator error mitigation by measuring the exchange rate to the spectator qubit as a function of coupler bias. A schematic of qubits 0, 1, and 4 is shown in Fig.~\ref{fig:gaps} (a). The corresponding level spectrum is shown in Fig.~\ref{fig:gaps} (b) as a function of the bias of the 1-4 coupler. At the point circled in Fig.~\ref{fig:gaps} (b) the frequency of qubit 0 crosses that of qubit 1. The net coupling $J_\text{Q0,Q1}$ between qubits 0 and 1 results in a gap with size $2J_\text{Q0,Q1}$ at the avoided crossing as shown in the inset of Fig.~\ref{fig:gaps} (b). 

For the case of the 1-4 gate with qubit 0 acting as a spectator, we can measure this exchange rate by preparing qubit 1 in the excited state and pulsing the 1-4 coupler to reach the center of the avoid crossing between qubits 0 and 1 and measuring the swap frequency. The resulting exchange rate $2J_\text{Q0,Q1}$ as a function of the DC bias being applied to the 0-1 coupler is shown in Fig.~\ref{fig:gaps} (c) for both states of qubit 4. We see that the bias which minimizes the exchange rate matches the optimal bias point for the 1-4 gate from Fig.~\ref{fig:rb} (a).  Furthermore when we repeat the experiment, but this time prepare Q4 in the excited state, we find slightly different cancellation value -- thus the cancellation bias is conditional on the state of Q4.

To determine the origin of these effects we model the device with the following Hamiltonian,

\begin{equation}
H_{\rm tot} = \sum H_{i} + H_{c}
\end{equation}
\begin{equation}
H_{i} = \hbar \omega_{i} a_i^{\dagger} a_i + \frac{\delta_i}{2} a^\dagger_i a^\dagger_i a_i a_i
\end{equation}
where $\omega_i$ is the frequency of the $i$-th transmon with, for the case of qubits 0-1-4, $i \in \{\mathrm{Q0},\mathrm{Q1},\mathrm{Q4},\mathrm{C01},\mathrm{C14}\} $, $\delta_i$ is the anharmonicity, and $a^\dagger_i$ ($a_i$) are the raising (lowering) operators.
The coupling Hamiltonian is given by
\begin{equation}
H_{\rm c} = \sum_{i \neq j} g_{i,j} \left( a^\dagger_i + a_i\right) \left( a^\dagger_j + a_j\right). 
\end{equation}

In an ideal device each qubit would only couple to the the neighboring tunable couplers, so that $g_{\text{Q0,C14}}$, $g_{
\text{Q4,C01}}$, and $g_{
\text{Q0,Q4}}$ would be zero. For this case we plot the predicted exchange rate as a function of flux as the solid blue and orange curves in Fig.~\ref{fig:gaps} (d) for the 0-1-4 qubits. As expected the exchange rate crosses zero at the same bias point for either state of Q4. This zero also matches approximately where we find the optimal bias point for simultaneous single-qubit operations on qubit 0 and 1. 

In real devices, however, there are finite microwave couplings between all elements. Specifically stray capacitive couplings like the one shown in Fig.~\ref{fig:gaps} (a) are the primary contributor to the spectator errors we observe. In Fig.~\ref{fig:gaps} (d) we model our device with stray couplings included as fit parameters and extract $g_{\text{Q0,C14}}$=3.7 MHz and $g_{\text{Q0,Q4}}$=80 kHz (blue and orange dashed curves), while from finite-element-method simulations of our full device structure we expect these couplings to be $g_{\text{Q0,C14}}$=1.5 MHz and $g_{\text{Q0,Q4}}$=80 kHz. We see that these additional coupling terms shift the zeros toward the region of high $ZZ$ and cause them to become conditional.

\begin{figure}%
	\centering
	\includegraphics[width=1\columnwidth]{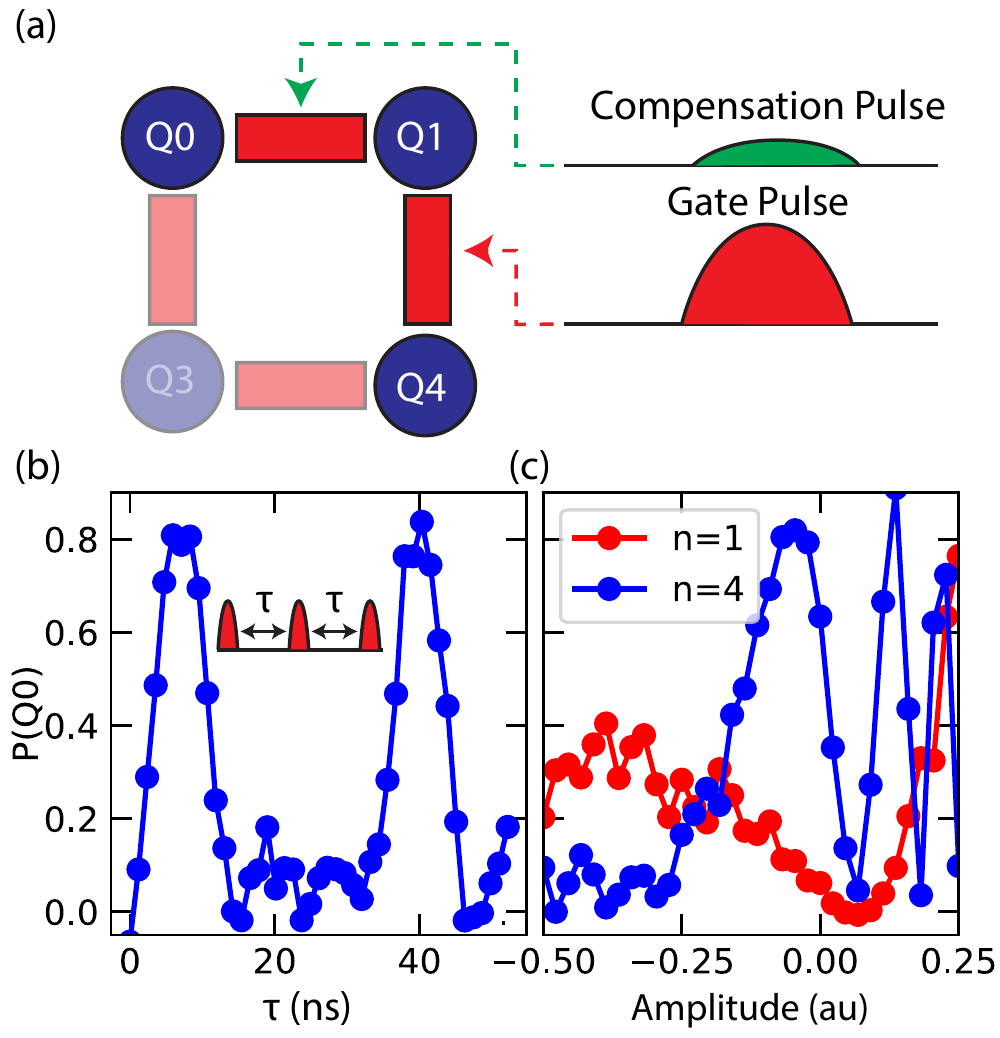}
	\caption{(a) Illustration of the compensation scheme.  During a gate between Q1 and Q4 we also apply a smaller pulse to C01, which is tuned up to cancel the partial swap. (b) Transfer probability as a function of the delay between gate pulses.  Constructive interference between population transfers of successive pulses is only achieved for specific delay times. (c) Population transfer with optimized delay time after $n$ pulses as a function of the amplitude of the compensation pulse.  A clear minimum is visible.}
	\label{fig:compcalib}
\end{figure}

\begin{figure}%
	\centering
	\includegraphics[width=1\columnwidth]{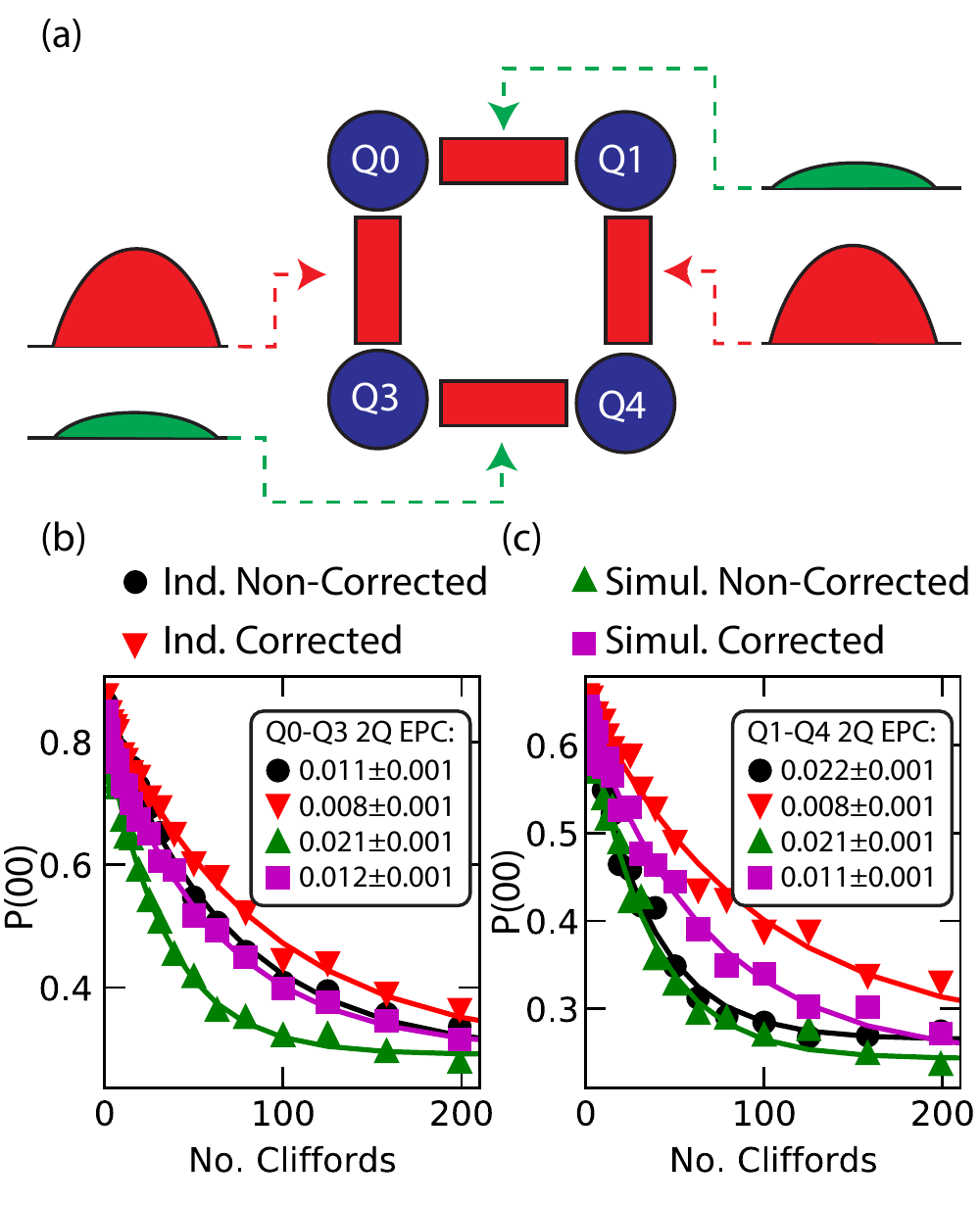}
	\caption{(a) Dynamic flux biasing scheme. While applying the primary flux pulses to the 0-3 and 1-4 coupler we apply small compensation pulses to the 3-4 and 0-1 couplers to minimize spectator errors throughout the course of each gate. Benchmarking data for the 0-3 and 1-4 gate are shown in (b) and (c) respectively. Individually measured benchmarking data with compensation(red) and without(black) as well as simultaneously measured benchmarking data with compensation(purple) and without(green) show that dynamically biasing to minimize spectator errors can drastically improve gate fidelities.}
	\label{fig:comp}
\end{figure}

Intuitively we can understand the conditional behaviour as an interference effect involving the new coupling pathway created by the stray capacitive coupling depicted in Fig.~\ref{fig:gaps} (a). During a CZ gate between Q1 and Q4 the weight of Q1's wavefunction in the C14 coupler increases significantly. Since C14 has a direct coupling to Q0, this turns on an interaction between Q1 and Q0. In order to balance this newly activated coupling pathway we have to bias the C01 coupler to an equal and opposite effective coupling strength. The dependence on the state of Q4 is a reflection of the fact that Q1 and Q4 strongly hybridize when C14 is in the on state, and therefore the weight of Q1 in the C14 coupler is dependent on Q4.

This intuitive picture lends itself to a cancellation scheme.  Since the spectator interaction is a partial swap it only occurs when the relevant two levels are degenerate.  We can therefore pulse the spectator bus such that when the collision occurs the relevant interaction is minimized.  This is schematically shown in Fig.~\ref{fig:compcalib}(a) -- when performing a gate between Q1 and Q4, we will also apply a compensation pulse to the C01 bus.   

To calibrate the amplitude of the compensation pulse, we need to measure small population transfers between Q1 and the spectator Q0.  We achieve this by measuring the population after several pulses.   Each pulse can be treated as a Landau-Zener transition (LZ) \cite{Oliver1653} during which part of the population of Q1 can swap into Q0, thus acting as a beam-splitter.  The series of pulses can then be thought of as an interferometer.   In general successive LZ transitions will not interfere constructively leading to only a slow population transfer.  As a result it is not enough to simply apply $n$ pulses, we also need to ensure the population transfers interfere constructively.  In our case we do this by applying four pulses and sweeping the delay $\tau$ between the pulses.  Since Q0 and Q1 have different frequencies changing the $\tau$ changes the accumulated phase difference between the swapped and non-swapped paths.  The result of this experiment is shown in Fig.~\ref{fig:compcalib}(b), where we plot the population of Q0 after 4 pulses applied to the C14 bus.  The plot features two prominent peaks that we identify with the condition for constructive interference. 

Fixing tau at the peak of constructive interference $\tau= 39$ ns we now sweep the amplitude of the correction pulse and measure population transfer from Q0 to Q1 for both one and four applied pulses.  This is shown in Fig.~\ref{fig:compcalib}(c).  The plot shows a minimum population transfer at an amplitude of 0.07.  
To realize high fidelity parallel gates we repeat this procedure for the Q0-Q1 pair, there applying the correction to the C14 bus.

We now extend this technique to parallel gates.  We focus on Q0, Q1, Q3 and Q4 qubits and drive the Q0-Q3 and Q1-Q4 gates. During these gates we apply a compensating pulses on C01 and C34 to dynamically bias the couplers, as shown in Fig.~\ref{fig:comp}(a), in a way that minimizes spectator errors between Q0-Q1 and Q3-Q4. We find that these compensation pulses create additional phase shifts on the qubits which can themselves produce large gate errors if not accounted for properly.   Fig.~\ref{fig:comp} (b,c) show the benchmarking results on the Q0-Q3 and Q1-Q4 gates respectively. The black curves show the fidelity we achieve individually benchmarking each gate, with the coupler to the spectator at the optimal bias for single-qubit gates. If we then individually benchmark the gates but add dynamic biasing to minimize spectator swaps, we get the fidelity shown by the red curves. Next, we benchmark the two gates simultaneously both without compensation pulses (green) and with (purple). While this technique can only account for the unconditional part of the shift in the cancellation bias, it still offers a significant advantage over static flux biasing.  In principle, due to the fact that the collisions that drive swaps in the 1 and 2 photon manifolds tend to occur at different amplitudes of the pulse, it should be possible to cancel the spectator swaps even in the presence of conditionality.

In low detuning devices with high stray couplings such as the one studied here, the need for dynamic biasing can introduce significant overhead in device bring up. However, the fact that the largest error we see is a resonant swap lends itself to simple mitigation strategies. In practice we find that detunings of more than $50$ MHz between the qubits is sufficiently large that we do not need to apply correction pulses. Through the use of laser-annealing and allocation of multiple bands of qubit frequencies \cite{Hertzberg2020}, we can make such qubit collisions a sufficiently rare occurrence that only small, sparse groups of qubits will need to have compensation pulses tuned up.
 
In summary, we've shown how spectator errors arise in a tunable coupling architecture, identified the key device parameter that gives rise to these errors, and demonstrated a simple software based strategy for mitigating these effects. While we have specifically analyzed these errors in the context of a BBQ architecture, similar considerations apply to any interference based coupling architecture where the weight of qubits' wavefunctions are altered by the execution of gate operations. While eliminating these stray couplings in hardware would be ideal, we've also demonstrated that these errors can be greatly diminished through the use of dynamic flux biasing. Combining these two approaches provides a pathway toward large scale devices with fast, tunable coupling and greatly enhanced spectator error cancellation.

\bibliography{references}
\end{document}